\documentclass[twocolumn,showpacs,preprintnumbers,amsmath,amssymb]{revtex4}

\usepackage{times}
\usepackage{natbib}
\usepackage{amssymb,amsbsy,amsmath,amsfonts}
\usepackage{graphicx}
\usepackage{float}
\usepackage{rotating}
\begin{document}
\title {The lowest-lying baryon masses in covariant SU(3)-flavor chiral perturbation theory}
\author{J. Martin Camalich$^1$}
\author{L.S. Geng$^{2,3,1}$}
\author{M.J. Vicente Vacas$^1$}
\affiliation{$^1$Departamento de F\'{\i}sica Te\'orica and IFIC, Universidad de
Valencia-CSIC, E-46071 Spain\\
$^2$School of Physics and Nuclear Energy Engineering, Beihang University,  Beijing 100191,  China\\
$^3$Physik Department, Technische Universit\" at M\"unchen, D-85747 Garching, Germany}
\begin{abstract}
We present an analysis of the baryon-octet and -decuplet masses using covariant SU(3)-flavor chiral perturbation theory up to next-to-leading order. Besides the description of the physical masses we address the problem of the lattice QCD extrapolation. Using the PACS-CS collaboration data we show that a good description of the lattice points can be achieved at next-to-leading order with the covariant 
loop amplitudes and phenomenologically determined values
for the meson-baryon couplings. Moreover, the extrapolation to the physical point up to this order is found to be better
than the linear one given at leading-order by the Gell-Mann-Okubo approach. The importance that a reliable combination of lattice QCD and 
chiral perturbation theory may have for hadron phenomenology is emphasized with the prediction of the pion-baryon and strange-baryon sigma terms.

\end{abstract}
\date{\today}
\maketitle

\section{Introduction}

The last few years have witnessed an impressive development of first-principles description of several hadronic observables 
by means of lattice QCD (lQCD)~\cite{Jansen:2008vs}. The studies of the lowest-lying baryon mass spectrum that have been 
undertaken by different collaborations using dynamical
actions with light quark masses close to the physical point~\cite{WalkerLoud:2008bp,Aoki:2008sm,Lin:2008pr,Durr:2008zz,Alexandrou:2009qu,Aoki:2009ix,Bietenholz:2010jr} are a remarkable example of this progress.
The quark-mass dependence of the baryon masses provides information on their scalar structure, i.e. sigma terms. The importance that a model-independent determination of these quantities have for dark matter searches has been recently pointed out~\cite{Ellis&Young}. 

Chiral perturbation theory ($\chi$PT), as the effective field theory of QCD at low-energies~\cite{ChiralIntro,Gasser:1987rb} is 
a suitable tool to analyze the lQCD results.
In principle, adjusting the proper low-energy constants (LECs) one shall optimally reproduce the quark mass dependence of the lQCD results 
and extrapolate them to the physical point.
$\chi$PT also provides a framework to ascertain systematic errors like those given 
by the boundaries of the finite box in which the simulation is performed. However, a full understanding of current $2+1$-flavor lQCD results using SU(3)-$\chi$PT is not yet available. A neat example
in the baryon sector (B$\chi$PT) is the difficulties found to describe the quark-mass dependence of the octet- and decuplet-baryon masses~\cite{WalkerLoud:2008bp,Ishikawa:2009vc} 
within the heavy-baryon (HB)$\chi$PT approach~\cite{Jenkins:1990jv}. Different solutions, like the SU(2)-flavor hyperon B$\chi$PT~\cite{SU2extr} or the use of different variants of cut-off renormalization~\cite{Donoghue:1998bs,Young:2009zb}, have been proposed 

Besides that,  in the last decades there have been a sustained interest in the SU(3)-B$\chi$PT description of the baryon-masses~\cite{ChPTMasses}. Recently, it has been shown that a successful study of hyperon phenomenology can be achieved using the covariant approach of
SU(3)-B$\chi$PT~\cite{Camalich:2009ax}. Examples include the description of the baryon-octet magnetic
moments~\cite{Geng:2008mf,Geng:2009hh} the analysis of the semileptonic hyperon decays~\cite{Geng:2009ik,Camalich:2009ax} and the prediction of the magnetic dipole moments and electromagnetic
structure of the decuplet resonances~\cite{Geng:2009ys}. The power counting is restored~\cite{Gasser:1987rb} using the so-called extended-on-mass-shell 
(EOMS) renormalization scheme~\cite{Fuchs:2003qc} and the contributions of decuplet resonances are systematically 
and explicitly considered ~\cite{Pascalutsa:2000kd,Geng:2009hh}. A successful extension of this formalism to the analysis 
of the lQCD results may lead to important and far-reaching phenomenological applications.

The purpose of the present work is to explore the potential of the covariant formulation of SU(3)-B$\chi$PT to conciliate the description of hyperon
phenomenology and the description of the quark mass dependence of $2+1$-flavor lQCD results, studying those of the PACS-CS collaboration on the baryon-octet and
-decuplet masses~\cite{Aoki:2008sm}. We have chosen the PACS-CS results because they contain simulations around the physical point and with different strange-quark masses. This gives us an opportunity to study the extrapolation in both
the pion and kaon masses. Moreover, the fine-grained lattices used by this collaboration, $a=0.0907(13)$ fm, allow to neglect the discretization 
errors in a first instance.  

\section{Formalism}

At $\mathcal{O}(p^2)$ the following terms in the chiral Lagrangian contribute to the octet and decuplet masses
\begin{eqnarray}
&&\mathcal{L}_{B}^{(2)}=b_0 \langle \chi_+\rangle\langle\bar{B}B\rangle+b_{D/F}\langle \bar{B}[\chi_{+},B]_\pm\rangle, \label{Eq:OctetCT}\\
&&\mathcal{L}_{T}^{(2)}=\frac{t_0}{2}\bar{T}^{abc}_\mu g^{\mu\nu}T_\nu^{abc}\langle\chi_{+}\rangle + \frac{t_D}{2}\bar{T}^{abc}_\mu g^{\mu\nu}\left(\chi_+,T_\nu\right)^{abc}, \nonumber
\end{eqnarray}
where $\langle X\rangle$ is the trace in flavor space and $(X,T_\mu)^{abc}\equiv(X)_d^a T_\mu^{dbc}+(X)_d^b T_\mu^{adc}+(X)_d^c T_\mu^{abd}$. 
In the Eqs. (\ref{Eq:OctetCT}), $\chi_+$ introduces the explicit chiral symmetry breaking, and
the coefficients $b_0$, $b_D$, $b_F$, and $t_0$, $t_D$ are unknown LECs. 

For the calculation of the leading loop contributions to the masses with pseudoscalar 
mesons ($\phi$), octet- ($B$) and decuplet-baryons ($T$) we use
the lowest-order $\phi B$ Lagrangian and the $\phi BT$ and $\phi T$ ones of Refs.~\cite{Geng:2009hh,Geng:2009ys}. 
The Lagrangians involving decuplet-baryons are $consistent$ in the sense that they constrain the unphysical degrees of freedom contained 
in the relativistic spin-3/2 fields~\cite{Pascalutsa:2000kd}. We take the values $D=0.80$ and $F=0.46$ for the $\phi B$ couplings, and
$\mathcal{C}=1.0$~\cite{Geng:2009hh} and  $\mathcal{H}=1.13$~\cite{Geng:2009ys} for the  $\phi B T$ and $\phi T$ ones. Notice that 
due to the factor 2 of difference between the HB and covariant conventions on $\mathcal{C}$ and $\mathcal{H}$~\cite{Geng:2009hh,Geng:2009ys}, our value for $\mathcal{C}$ is 
larger than the one
often used in the HB literature~\cite{ChPTMasses}. For the meson decay constant we take an average $F_\phi\equiv1.17f_\pi$ 
with $f_\pi=92.4$ MeV. For the masses of the pseudoscalar mesons we use $m_\pi\equiv m_{\pi^\pm}=139$ MeV, $m_K\equiv m_{K^\pm}=494$ MeV. 
The mass of the $\eta$ meson is fixed with the Gell-Mann-Okubo mass relation, $3m_\eta^2=4m_K^2-m_\pi^2$.

At $\mathcal{O}(p^2)$, the Lagrangians in Eqs. (\ref{Eq:OctetCT}) give a tree-level contribution and provide the leading SU(3)-breaking 
to the chiral limit masses $M_{B0}$ and $M_{D0}$ appearing in the free Lagrangians. The breaking is linear in the quark
masses and leads to the Gell-Mann-Okubo mass relation for the octet, $3M_\Lambda+M_\Sigma-2(M_N+M_\Xi)=0$, and the Gell-Mann decuplet equal
spacing rules, $M_{\Sigma^*}-M_\Delta=M_{\Xi^*}-M_{\Sigma^*}=M_{\Omega^-}-M_{\Xi^*}$~\cite{ChPTMasses}. In the following, 
we refer to these results that we recover in B$\chi$PT at the leading order (LO) calculation of the self-energy as the 
Gell-Mann-Okubo (GMO) results.

The next-to-leading order (NLO) contribution to the self-energy of the baryons is given at  
$\mathcal{O}(p^3)$ by the leading loop diagrams~\cite{ChPTMasses}. These only depend on phenomenologically determined couplings and masses and no new unknown LECs are introduced at this order. 
We calculate the loops in the covariant approach and recover the power-counting using the EOMS renormalization 
prescription introduced in Ref.~\cite{Fuchs:2003qc} and generalized for the decuplet loops in Ref.~\cite{Geng:2009hh}. From the covariant 
results we have obtained the HB ones by defining $M_D=M_B+\delta$ and expanding the loop-functions about the limit 
$M_B\rightarrow \infty$~\cite{ChPTMasses}. A full presentation of the NLO results used in this work can be found in the Appendix. 
 
\begin{table}
\centering
\caption{Results of the fits to the physical values of the baryon-octet masses (in MeV) in B$\chi$PT up to NLO.
\label{Table:ResMassExp}}
\begin{ruledtabular}
\begin{tabular}{ccccc}
&GMO&HB&Covariant&Expt.\\
\hline
$M_N$&$942(2)$&$939(2)$&$941(2)$&$940(2)$\\
$M_\Lambda$&$1115(1)$&$1116(1)$&$1116(1)$&$1116(1)$\\
$M_\Sigma$&$1188(4)$&$1195(4)$&$1190(4)$&$1193(5)$\\
$M_\Xi$&$1325(3)$&$1315(3)$&$1322(3)$&$1318(4)$\\
\hline
$M_{B0}^{eff}$ [MeV] &1192(5)&2422(5)&1840(5)&\\
$b_D$ [GeV$^{-1}$]&0.060(4)&0.412(4)&0.199(4)&\\
$b_F$ [GeV$^{-1}$]&$-$0.213(2)&$-$0.781(2)&$-$0.530(2)&\\
\end{tabular}
\end{ruledtabular}
\end{table}

\section{Results}
\subsection{Analysis of experimental data}

In Table~\ref{Table:ResMassExp} we present the results of fits to the physical values of the 
baryon-octet masses in B$\chi$PT up to $\mathcal{O}(p^3)$. We assign an error bar to the experimental values that accounts for the mass splitting within any isospin multiplet with a minimum error of 1 MeV for the isospin singlets. The term associated to $b_0$ in Eqs. (\ref{Eq:OctetCT}) 
cannot be disentangled from $M_{B0}$  when comparing to the experimental data~\cite{ChPTMasses}. Therefore, we define
$M_{B0}^{eff}=M_{B0}-b_0(4m_K^2+2m_\pi^2)$ and perform the fits  using 
$M_{B0}^{eff}$, $b_D$ and $b_F$ as free parameters. Equivalently, we present in Table~\ref{Table:ResMassDecExp} the B$\chi$PT results 
for the decuplet-baryons with $M_{D0}^{eff}=M_{D0}-t_0(2m_K^2+m_\pi^2)$ and $t_D$ as the fitting parameters. The 
octet-decuplet mass splitting is fixed to the separation between the averaged physical masses, $\delta=231$ MeV, in the diagrams with baryons of both multiplets.

As we have seen above, the starting point for a description of the SU(3)-breaking expansion of the lowest-lying baryon masses in 
B$\chi$PT is the GMO relations which are known to work with an accuracy of $\sim7$ MeV. This is reflected by 
the good numerical results of the fit shown in the first column of Tables~\ref{Table:ResMassExp} and~\ref{Table:ResMassDecExp}.
A puzzling and well known feature of the leading chiral loops is that they preserve the GMO equations within $\sim10$ MeV whereas the corrections given to any of the masses ($\delta M_{B/T}$) are of order $\sim$100-1000 MeV~\cite{ChPTMasses,Jenkins:2009wv}. Indeed, 
in Table~\ref{Table:ResMassExp} we see that at NLO the description for the octet is improved few MeV despite large higher-order
contributions to the masses, i.e. $\delta M_N=-228\;{\rm MeV}$ or $\delta M_\Xi=-799\;{\rm MeV}$ 
in the covariant approach (in HB the corrections are almost doubled). In the case of the decuplet resonances (Table~\ref{Table:ResMassDecExp}) at NLO the description in the covariant approach
is reasonably good, despite again the large chiral corrections to the masses, i.e. $\delta M_\Delta=-324$ or $\delta M_{\Omega^-}=-821\;{\rm MeV}$.

\begin{table}
\centering
\caption{Results of the fits to the physical values of the baryon-decuplet masses (in MeV) in B$\chi$PT up to NLO.  \label{Table:ResMassDecExp}}
\begin{ruledtabular}
\begin{tabular}{ccccc}
&GMO&HB&Covariant&Expt.\\
\hline
$M_\Delta$&$1233(2)$&$1235(2)$&$1234(2)$&$1232(2)$\\
$M_{\Sigma^*}$&$1380(1)$&$1372(1)$&$1376(1)$&$1385(4)$\\
$M_{\Xi^*}$&$1526(1)$&$1518(1)$&$1523(1)$&$1533(4)$\\
$M_{\Omega^-}$&$1672(1)$&$1673(1)$&$1673(1)$&$1672(1)$\\
\hline
$M_{D0}^{eff}$ [MeV] &1215(2)&1763(2)&1519(2)\\
$t_D$ [GeV$^{-1}$]&$-0.326(2)$&$-0.960(2)$&$-0.694(2)$\\
\end{tabular}
\end{ruledtabular}
\end{table}

\subsection{Analysis of lQCD results}

For the study of the lQCD results, the convergence properties of the different approaches to $\chi$PT are more critical because of the larger quark mass involved in the simulations. In the following, we extend our analysis to the 2+1-flavor lQCD results, in particular to those reported by the 
PACS-CS collaboration in Ref.~\cite{Aoki:2008sm}. We choose the points for which both the pion and kaon masses are below 
600 MeV which is a limit of the meson mass we deem acceptable for the convergence of covariant SU(3)-B$\chi$PT up to NLO~\cite{Geng:2008mf,Geng:2009hh}. 
Namely, we take the lightest three lattice points, including one with a lighter strange quark mass, for each baryon. The 
masses in physical units are obtained using the lattice spacing $a=0.0907(13)$ fm, where $a$ is determined using the physical values of $m_\pi$, $m_K$ and $M_{\Omega^-}$~\cite{Aoki:2008sm}. 

The analysis of the lQCD results disentangles the singlet parts $b_0$ and $t_0$ from the respective chiral limit masses. 
Therefore we perform fits of the 7 parameters, $M_{B0}$, $b_0$, $b_D$, $b_F$, $M_{D0}$, $t_0$ and $t_D$, to the chosen 
24 lattice points that we assume to have independent statistical errors ($\sigma_i$) but fully-correlated errors propagated from $a$ ($\Delta a_i$). Our $\chi^2$ incorporates the inverse of the resulting $24\times 24$ correlation matrix $C_{ij}=\sigma_i\sigma_j\delta_{ij}+\Delta a_i\Delta a_j$. Besides, the fits to the octet and decuplet masses are connected through the diagrams with baryons of both 
multiplets. Finally, the 
finite volume corrections have been calculated in the covariant approach and included to the analysis~\cite{Beane:2004tw}.
The largest corrections are found for $M_N$ ($-7$ MeV) in the octet and for $M_\Delta$ ($-25$ MeV) in the decuplet at the lightest lattice point, for which $m_\pi L\simeq2<4$~\cite{Aoki:2008sm}. Thus there could be further finite volume uncertainties at this point, although we do not expect them to have a big influence in the fits given the large statistical error bars assigned to the corresponding simulated baryon masses.    

\begin{table}
\centering
\caption{Extrapolation in MeV and values of the LECs from the fits to the PACS-CS results~\cite{Aoki:2008sm} on the baryon masses using B$\chi$PT up to
NLO. The $\chi^2$ is the estimator for the fits to the lQCD results whereas $\bar{\chi}^2$ include also experimental data. See the text for details. \label{Table:ResChExtrapolation}}
\begin{ruledtabular}
\begin{tabular}{ccccc}
&GMO&HB&Covariant&Expt.\\
\hline
$M_N$&$971(22)$&$764(21)$&$893(19)(39)$&$940(2)$\\
$M_\Lambda$&$1115(21)$&$1042(20)$&$1088(20)(14)$&$1116(1)$\\
$M_\Sigma$&$1165(23)$&$1210(22)$&$1178(24)(7)$&$1193(5)$\\
$M_\Xi$&$1283(22)$&$1392(21)$&$1322(24)(20)$&$1318(4)$\\

$M_\Delta$&$1319(28)$&$1264(22)$&$1222(24)(49)$&$1232(2)$\\
$M_{\Sigma^*}$&$1433(27)$&$1466(22)$&$1376(24)(29)$&$1385(4)$\\
$M_{\Xi^*}$&$1547(27)$&$1622(23)$&$1531(25)(8)$&$1533(4)$\\
$M_{\Omega^-}$&$1661(27)$&$1733(25)$&$1686(28)(13)$&$1672(1)$\\
\hline
$M_{B0}$ [MeV] &900(39) &508(32)&756(32) \\
$b_0$ [GeV$^{-1}$]&$-0.264(24)$&$-1.656(19)$&$-0.978(38)$ \\
$b_D$ [GeV$^{-1}$]& 0.042(9) &$0.368(9)$&0.190(24) \\
$b_F$ [GeV$^{-1}$]& $-0.174(7)$ &$-0.824(6)$&$-0.519(19)$ \\
\hline
$M_{D0}$ [MeV]&1245(48) &1117(32)&954(37) \\
$t_0$ [GeV$^{-1}$]&$-0.12(5)$ &$-0.709(37)$&$-1.05(8)$ \\
$t_D$ [GeV$^{-1}$]&$-0.254(10)$ &$-0.739(10)$&$-0.682(20)$ \\
\hline
$\chi^2_{\rm d.o.f.}$& $0.63$ &$9.2$&$2.1$ \\
$\bar{\chi}^2_{\rm d.o.f.}$& $4.2$ &$36.6$&$2.8$ \\
\end{tabular}
\end{ruledtabular}
\end{table}

The results of the fits  are shown in Table~\ref{Table:ResChExtrapolation}. The error bar quoted in the GMO and HB columns and the first one assigned to the covariant results are the uncertainties propagated from the fitted parameters. The second error bar in the covariant results is a theoretical uncertainty coming from the truncation of the chiral expansion which is estimated by taking 1/2 of the difference between the results obtained at LO and NLO. 

At LO (GMO) the description of the lattice points has a $\chi^2_{\rm d.o.f.}$ that is smaller than one and the values of the masses 
extrapolated to the physical point are quite good. The extrapolation is expected to improve with the addition 
of the leading non-analytic chiral terms at NLO.
However, in the HB approach the description of the quark-mass dependence is much worse and the extrapolated values are $\sim$ 100 
MeV off the experimental ones. This is consistent with the problems reported in Refs.~\cite{WalkerLoud:2008bp,Ishikawa:2009vc}
where reasonable fits using SU(3)-HB$\chi$PT are only obtained with unreasonably small values (even compatible with zero) of the 
$\phi B$ and $\phi BT$ couplings.
\begin{table*}
\centering
\caption{Predictions on the $\sigma_\pi$ and $\sigma_s$ terms (in MeV) in covariant SU(3)-B$\chi$PT after fitting the 
LECs to the PACS-CS results.   \label{Table:ResSigmas}}
\begin{ruledtabular}
\begin{tabular}{ccccccccc}
&$N$&$\Lambda$&$\Sigma$&$\Xi$&$\Delta$&${\Sigma^*}$&${\Xi^*}$&${\Omega^-}$\\
\hline
$\sigma_\pi$&59(2)(17)&39(1)(10)&26(2)(5)&13(2)(1)&55(4)(18)&39(3)(13)&22(3)(7)&5(2)(1)\\
$\sigma_s$&$-4$(23)(25)&126(26)(35)&159(27)(45)&267(31)(50)&56(24)(1)&160(28)(7)&274(32)(9)&360(34)(26)\\
\end{tabular}
\end{ruledtabular}
\end{table*}

\begin{figure}[t]
\includegraphics[width=\columnwidth]{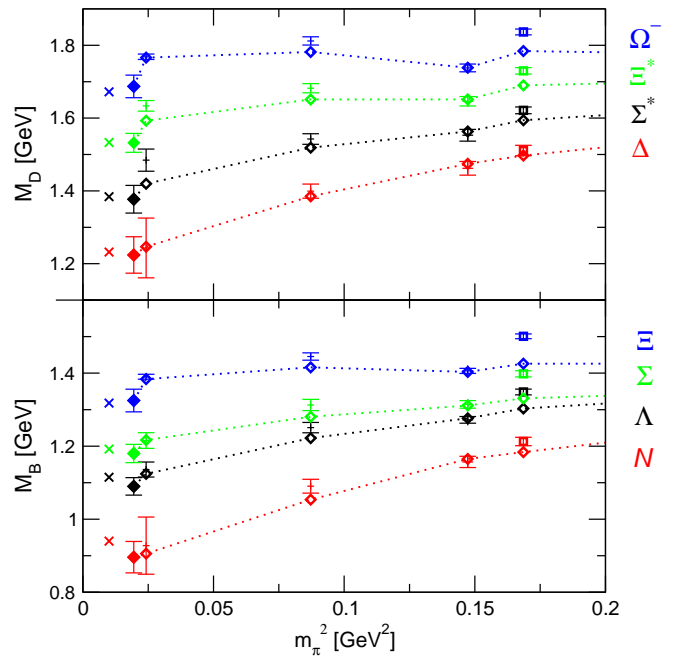}
\caption{(Color on-line) Extrapolation of the PACS-CS results~\cite{Aoki:2008sm} on the lowest-lying baryon masses within the 
covariant formulation of SU(3)-B$\chi$PT up to NLO. The lQCD points used in the 
fit are represented with the corresponding error bars which do not include the correlated uncertainties. The lattice points in $m_\pi^2\simeq0.15$ GeV$^2$ involve a lighter strange quark mass. The diamonds denote our  
results after the fit and they are connected by a dotted line added to guide the eye.
The boxes are lattice points not included in the fit (heavier kaon mass) and the filled diamonds are the extrapolated values which are to be compared with experimental data (crosses). The latter are slightly shifted for a better comparison with the extrapolation results. \label{fig_graph}}
\end{figure}

On the other hand, in the covariant formulation of SU(3)-B$\chi$PT, a good description of the lQCD results can be achieved using the phenomenological values for these couplings. Moreover, the comparison between the extrapolated values in the covariant and GMO frameworks suggests the importance of the non-analytic chiral structure nearby the physical point. This can be more quantitatively assessed including the experimental data into the fits. Their quality is shown in Table~\ref{Table:ResChExtrapolation} through the corresponding estimator that we denote by $\bar{\chi}^2_{\rm d.o.f.}$. We indeed see that the covariant approach accommodates the PACS-CS results and experimental data manifestly better than the GMO based model. This improvement, obtained at
NLO in covariant SU(3)-B$\chi$PT, highlights the effect of the leading 
non-analytical terms in the extrapolation even from light quark masses as small as those used in Ref.~\cite{Aoki:2008sm} ($m_\pi\simeq156$ MeV). We have also studied the role of the meson-baryon couplings by treating them as free parameters in the fit. The fitted values turn out to be consistent with the phenomenological ones, albeit with relatively large uncertainties.

In Fig.~\ref{fig_graph}, we show the quark-mass dependence of the baryon masses in covariant SU(3)-B$\chi$PT compared to
the PACS-CS results. As we 
can see in the figure, the light and strange quark mass dependence of the PACS-CS results on the masses are very well described in covariant SU(3)-B$\chi$PT. The comparison
of the heavy kaon points (boxes) with our results indicates that the agreement between the PACS-CS calculation and the SU(3)-B$\chi$PT NLO amplitude is
still quite good at these relatively large quark masses. 

\subsection{Determination of LECs and baryonic $\sigma$-terms}

The last issue we address in this work concerns the determination of the LECs of SU(3)-B$\chi$PT using 2+1-flavor 
simulations. Namely, we can see  in Table~\ref{Table:ResChExtrapolation} that the values of $b_D$, $b_F$ and $t_D$ obtained 
in the covariant approach closely agree with those shown in Tables~\ref{Table:ResMassExp} and 
\ref{Table:ResMassDecExp} which are obtained fitting experimental data. The lQCD results also allow to fix $M_{B0}$ and $M_{D0}$ 
separately from $b_0$ and $t_0$
and to study the convergence. Indeed, we have checked that the overall NLO loop contributions to the baryon masses are about 50$\%$-60$\%$ those
obtained at LO, which is consistent with the convergence expected for the SU(3)-flavor chiral expansion.

Nevertheless, the most relevant consequence of fixing the LECs within a reliable combination of lQCD and $\chi$PT is that 
it may lead to solid predictions. This can be illustrated in the scalar
sector with the determination of the sigma terms from the analysis of the masses through the Hellman-Feynman theorem~\cite{ChPTMasses,Young:2009zb}.
In Table ~\ref{Table:ResSigmas} we present the results on $\sigma_\pi$ and $\sigma_s$ after fitting the LECs to the 
PACS-CS results and with the uncertainties determined as has been discussed above for the masses. It is interesting to note that our results for the baryon octet are in agreement with those of Ref.~\cite{Young:2009zb} obtained within the cut-off renormalized B$\chi$PT. 

\section{Conclusions}

In summary, we have explored the applicability of the covariant formulation of SU(3)-B$\chi$PT within the EOMS scheme to analyze current 2+1-flavor
lQCD data, i.e. the results of the PACS-CS on the baryon masses. In contrast with the problems found in HB, the 
covariant approach is able to describe simultaneously the experimental data and PACS-CS results. Moreover, we have found that the consistency between both
is improved from the good linear extrapolation obtained at LO (GMO) with the inclusion of the leading non-analytic terms. This is achieved despite that the small light quark masses from which we perform the extrapolation are very close to the physical point. The  
success of a SU(3)-B$\chi$PT approach to describe current 2+1-flavor lQCD results may have important phenomenological applications, as has been shown with the
determination of the $\sigma$ terms. Furthermore, it is necessary to stress that the present study has been based in one lattice spacing. Discretization uncertainties may still be sizable for some of the baryons and these should be
studied more carefully in the future. An analysis of lQCD results obtained by other collaborations is in progress. 
  
\section{Acknowledgments}

We want to thank L. Alvarez-Ruso, J.A. Oller and S. Scherer for independently motivating the present study. Discussions with J. Nieves and R.D. Young
are acknowledged. This work was partially supported by the  MEC 
contract  FIS2006-03438 and the EU Integrated Infrastructure
Initiative Hadron Physics Project contract RII3-CT-2004-506078. JMC acknowledges the MICINN for support. LSG's work was partially supported by the Fundamental Research Funds for the Central Universities, the Alexander von Humboldt foundation, and the MICINN in the program Juan de la Cierva.

\section{Appendix}

\begin{figure}[t]
\includegraphics[width=8cm]{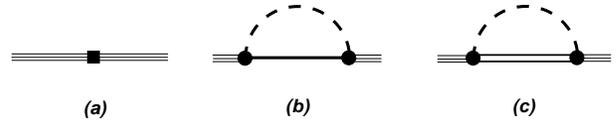}
\caption{Feynman diagrams contributing to the octet- and decuplet-baryons ($B$ and $T$ respectively) up to 
$\mathcal{O}(p^3)$ in $\chi$PT. The solid lines correspond to octet-baryons, double lines to decuplet-baryons and dashed 
lines to mesons. The black dots indicate $1^{st}$-order couplings while boxes, $2^{nd}$-order couplings. 
\label{Fig:masses}}
\end{figure}

In Figure ~\ref{Fig:masses} we show the Feynman diagrams that contribute to the self-energy of the lowest-lying baryons in $\chi$PT up to $\mathcal{O}(p^3)$ or NLO. At $\mathcal{O}(p^2)$ the tree-level contribution $(a)$ of Fig.~\ref{Fig:masses} introduces the LECs, $b_0$, $b_D$, $b_F$, $t_0$ and $t_D$, that provide the LO SU(3)-breaking corrections to the chiral limit masses appearing in the free Lagrangians
\begin{equation}
M_\mathcal{B}^{(2)}=M_0-\sum_{\phi=\pi,K}\xi^{(a)}_{\mathcal{B},\phi}m_\phi^2,\label{Eq:MassesTreeLevel}
\end{equation}
where  we are explictly using the Gell-Mann-Okubo (GMO) equation to relate the $\eta$, $\pi$ and $K$ masses. With $\mathcal{B}$ we denote both octet- and decuplet-baryons $\mathcal{B}=N$, $\Lambda$, $\Sigma$, $\Xi$ or $\Delta$, $\Sigma^*$, $\Xi^*$, $\Omega^-$ and $M_0=M_{B0}$ or $M_{D0}$ respectively. The coefficients $\xi^{(a)}_{\mathcal{B},M}$ are SU(3) Clebsch-Gordan coefficients that can be found in Table~\ref{Table:Coefficients}. 

\begin{table*}
\caption{Coefficients of the loop-contribution to the self-energy Eq.~(\ref{Eq:ResGen}) for any of the octet or decuplet baryons. \label{Table:Coefficients}}
\hspace{-1cm}
\begin{tabular}{ccccccccc}
\hline
&$N$&$\Lambda$&$\Sigma$&$\Xi$&$\Delta$&$\Sigma^*$&$\Xi^*$&$\Omega^-$\\
\hline
$\xi_{\mathcal{B},\pi}^{(a)}$&$2 b_0+4 b_F$&$\frac{2}{3} \left(3 b_0-2 b_D\right)$&$2 b_0+4 b_D$&$2 b_0-4 b_F$&$ t _0+3  t_D$&$ t _0+ t_D$&$ t _0- t_D$&$  t _0-3  t_D$\\
$\xi_{\mathcal{B},K}^{(a)}$&$ 4 b_0+4 b_D-4 b_F$&$\frac{2}{3} \left(6 b_0+8 b_D\right)$&$4 b_0$&$4 b_0+4 b_D+4 b_F$&$2t_0$&$2t _0+ 2t_D$&$ 2t _0+4 t_D$&$ 2 t _0+6  t_D$\\
$\xi_{\mathcal{B},\pi}^{(b)}$&$\frac{3}{2}(D+F)^2$&$2D^2$&$\frac{2}{3} (D^2+6F^2)$&$\frac{3}{2} (D-F)^2$&$\frac{4}{3}\mathcal{C}^2$&$\frac{10}{9}\mathcal{C}^2$&$\frac{2}{3}\mathcal{C}^2$&$0$\\
$\xi_{\mathcal{B},K}^{(b)}$&$\frac{1}{3}(5D^2-6DF+9F^2)$&$\frac{2}{3}(D^2+9F^2)$&$2(D^2+F^2)$&$ \frac{1}{3} (5D^2+6DF+9F^2)$&$\frac{4}{3}\mathcal{C}^2$&$\frac{8}{9}\mathcal{C}^2$&$\frac{4}{3}\mathcal{C}^2$&$\frac{8}{3}\mathcal{C}^2$\\
$\xi_{\mathcal{B},\eta}^{(b)}$&$\frac{1}{6}(D-3F)^2$&$ \frac{2}{3} D^2$&$\frac{2}{3} D^2$&$\frac{1}{6} (D+3F)^2$&$0$&$\frac{2}{3}\mathcal{C}^2$&$\frac{2}{3}\mathcal{C}^2$&$0$\\
$\xi_{\mathcal{B},\pi}^{(c)}$&$\frac{16}{3}\mathcal{C}^2$&$4\mathcal{C}^2$&$\frac{8}{9}\mathcal{C}^2$&$\frac{4}{3}\mathcal{C}^2$&$\frac{50}{27}\mathcal{H}^2$&$\frac{80}{81}\mathcal{H}^2$&$\frac{10}{27}\mathcal{H}^2$&$0$\\
$\xi_{\mathcal{B},K}^{(c)}$&$\frac{4}{3}\mathcal{C}^2$&$\frac{8}{3}\mathcal{C}^2$&$\frac{40}{9}\mathcal{C}^2$&$4\mathcal{C}^2$&$\frac{20}{27}\mathcal{H}^2$&$\frac{160}{81}\mathcal{H}^2$&$\frac{20}{9}\mathcal{H}^2$&$\frac{40}{27}\mathcal{H}^2$\\
$\xi_{\mathcal{B},\eta}^{(c)}$&$0$&$0$&$\frac{4}{3}\mathcal{C}^2$&$\frac{4}{3}\mathcal{C}^2$&$\frac{10}{27}\mathcal{H}^2$&$0$&$\frac{10}{27}\mathcal{H}^2$&$\frac{40}{27}\mathcal{H}^2$\\
\hline
\end{tabular}
\end{table*}

At $\mathcal{O}(p^3)$ the graphs $(b)$ and $(c)$ give the NLO SU(3)-breaking corrections to the baryon masses
\begin{equation}
M^{(3)}_{\mathcal{B}}=M^{(2)}_{\mathcal{B}}+\frac{1}{(4\pi F_\phi)^2}\sum_{\substack{\phi=\pi,K,\eta\\ \alpha=b,c}}\xi^{(\alpha)}_{\mathcal{B},\phi} H_X^{(\alpha)}(m_\phi).\label{Eq:ResGen}
\end{equation}
The coefficients $\xi^{(\alpha)}_{\mathcal{B},M}$ are again Clebsch-Gordan coefficients that can be found in Table~\ref{Table:Coefficients} and $H_X^{(\alpha)}(m)$ are loop functions which are different for octet ($X=B$) or decuplet ($X=T$) external baryon lines. The loop functions can be cast in a compact form in term of Feynman-parameter integrals
\begin{eqnarray}
&&H^{(b)}_B=-\frac{M_{B0}^3}{2} \int^1_0dx\,\left(\left( x^3+3 (x+1) \bar{\mathcal{M}}_B^{(b)}\right)\right.\nonumber\\ 
&&\left.\times\left(\lambda_\varepsilon+\log \left(\frac{ \bar{\mathcal{M}}_B^{(b)}}{\bar{\mu}_B ^2}\right)\right)-(2
   x+1)\bar{\mathcal{M}}_B^{(b)}\right),\label{Eq:HBb}\\
&&H^{(c)}_B=-\frac{3M_{B0}^3}{4}\left(\frac{1}{R}\right)^2\int^1_0dx\, \left((1-x)+R\right)\bar{\mathcal{M}}_B^{(c)}\nonumber\\ 
&&\times \left(\lambda_\varepsilon+\log \left(\frac{\bar{\mathcal{M}}_B^{(c)}}{\bar{\mu}_B ^2}\right)\right),\label{Eq:HBc}\\ 
&&H^{(b)}_T=-\frac{3M_{D0}^3}{4}\int^1_0dx\, \left(r+(1-x)\right)\bar{\mathcal{M}}_T^{(b)} \nonumber\\ 
&&\times\left(\lambda_\varepsilon+\log \left(\frac{\bar{\mathcal{M}}_T^{(b)}}{\bar{\mu}_T
   ^2}\right)-1\right),\label{Eq:HTb}
\end{eqnarray}
\begin{eqnarray}
 &&H^{(c)}_T=-\frac{M_{D0}^3}{20} \int^1_0dx\, (2-x)\bar{\mathcal{M}}_T^{(c)} \nonumber \\
&&\times\left(15\left(\lambda_\varepsilon+\log \left(\frac{\bar{\mathcal{M}}_T^{(c)}}{\bar{\mu}_T ^2}\right)\right)+11\right). \label{Eq:HTc}
\end{eqnarray}
In the last equations we use the dimensionless quantities $R=M_{D0}/M_{B0}$ and $r=1/R=M_{B0}/M_{D0}$ for the baryon masses and $\bar{\mu}_B=\mu/M_{B0}$ or $\bar{\mu}_T=\mu/M_{D0}$ for the renormalization scale. Moreover, $\bar{\mathcal{M}}_{B/T}^{(\alpha)}=\mathcal{M}_{B/T}^{(\alpha)}/M_{B0/D0}$, where $\alpha=b$ or $c$ and 
\begin{eqnarray}
\mathcal{M}_{B/T}^{(b)}=(1-x)m^2+x\,M_{B0}^2-x(1-x)M_{B0/D0}^2-i\epsilon,\nonumber\\
\mathcal{M}_{B/T}^{(c)}=(1-x)m^2+x\,M_{D0}^2-x(1-x)M_{B0/D0}^2-i\epsilon,\nonumber
\end{eqnarray}
with $m$ the mass of the meson $\phi$ in the loop. 

The loop-integrals in Eqs. (\ref{Eq:HBb})-(\ref{Eq:HTc}) contain the divergent piece $\lambda_\varepsilon=2\mu^{\varepsilon}/\varepsilon+\log{4\pi}-\gamma_E$ ($d=4-\varepsilon$) and we cancel it renormalizing the LECs in the $\overline{MS}$-scheme. The resulting loop-functions still contain analytical finite pieces that break the power counting. We absorb them into the LECs within an extension of the $\overline{MS}$-scheme called EOMS-renormalization scheme~\cite{Fuchs:2003qc,Geng:2009hh}. This is equivalent to redefining the $\overline{MS}$-renormalized LECs as
\begin{eqnarray}
&&\tilde{M}_{B0}^r(\mu)=M^r_{B0}(\mu)\nonumber\\
&&+\frac{M_{B0}^3}{24\pi^2 F_\phi^2}\left[\left(5 D^2+9 F^2\right)\left(2\log \left(\bar{\mu}_B\right)+1\right)-5 \mathcal{C}^2f_0(\mu) \right],\nonumber
\end{eqnarray}
\begin{eqnarray}
&&\tilde{b}_0^r(\mu)=b_0^r(\mu)\nonumber\\
&&-\frac{M_{B0}}{144\pi^2 F_\phi^2}\left[\left(13 D^2+9 F^2\right)\left( \log\left(\bar{\mu}_B\right)+1\right)+\frac{7\mathcal{C}^2}{12}f(\mu)\right],\nonumber
\end{eqnarray}
\begin{eqnarray}
&&\tilde{b}_D^r(\mu)=b_D^r(\mu)\nonumber\\
&&+\frac{M_{B0}}{32\pi^2 F_\phi^2}\left[\left(D^2-3 F^2\right)\left( \log\left(\bar{\mu}_B\right)+1\right)+\frac{\mathcal{C}^2}{18}f(\mu)\right],\nonumber
\end{eqnarray}
\begin{eqnarray}
&&\tilde{b}_F^r(\mu)=b_F^r(\mu)\nonumber\\
&&-\frac{M_{B0}}{48\pi^2 F_\phi^2}\left[5 D F\left( \log\left(\bar{\mu}_B\right)+1\right)+\frac{5\mathcal{C}^2}{72}f(\mu)\right],\nonumber
\end{eqnarray}
\begin{eqnarray}
&&\tilde{M}_{D0}^r(\mu)=M^r_{D0}(\mu)\nonumber\\
&&+\frac{M_{D0}^3}{576\pi^2 F_\phi^2}\left[\frac{10 \mathcal{H}^2}{9} \left(60\log \left(\bar{\mu}_T\right)+1 \right)+\mathcal{C}^2g_0(\mu)\right],\nonumber
\end{eqnarray}
\begin{eqnarray}
&&\tilde{t}_0^r(\mu)=t_0^r(\mu)\nonumber\\
&&-\frac{M_{D0}}{576\pi^2 F_\phi^2}\left[\frac{25 \mathcal{H}^2}{27} (30 \log (\bar{\mu}_T )+23)+\mathcal{C}^2 g(\mu)\right],\nonumber
\end{eqnarray}
\begin{eqnarray}
&&\tilde{t}_D^r(\mu)=t_D^r(\mu)\nonumber\\
&&-\frac{M_{D0}}{1728\pi^2 F_\phi^2}\left[\frac{5 \mathcal{H}^2}{3} (30 \log (\bar{\mu}_T )+23)+\mathcal{C}^2 g(\mu)\right],
\end{eqnarray}
where the functions $f_0$, $f$, $g_0$ and $g$
\begin{eqnarray}
&&f_0(\mu)=\left(6 R^6+12 R^5-9 R^4-48 R^3-14 R^2+20 R\right.\nonumber\\
&&+\left.12 \left(1-2 R \left(3 R^2+R-1\right)\right) \log (\bar{\mu}_B )\right.\nonumber\\
&&\left.-12
   (R-1)^3 (R+1)^5 \log (R)\right.\nonumber\\
&&\left.+6 \left(2 R \left(3 R^2+R-1\right)-1\right) \log \left(R^2\right)\right.\nonumber\\
&&\left.+6 (R-1)^3 (R+1)^5 \log
   \left(R^2-1\right)+10\right)/(48R^2),\nonumber
\end{eqnarray}
\begin{eqnarray}
&&f(\mu)=\left(6 R^4+9 R^3+3 R^2+9 R+6 (3 R+2) \log (\bar{\mu}_B)\right.\nonumber\\
&&\left.-6 \left(2 R^6+3 R^5-3 R-2\right) \log (R)-3 (3 R+2) \log
   \left(R^2\right)\right.\nonumber\\
&&\left.+\left(6 R^6+9 R^5-9 R-6\right) \log \left(R^2-1\right)+7\right)/R^2,\nonumber\\
&&g_0(\mu)=-6 (r-1)^3 \log \left(\frac{1}{r^2}-1\right) (r+1)^5+\nonumber\\
&&r (r (3 r (r (3-2 r (r+2))+28)+26)-32)\nonumber\\
&&-12 \left(1-2 r
   \left(3 r^2+r-1\right)\right) \log \left(\frac{\bar{\mu}_T}{r}\right)-16,\nonumber\\
&&g(\mu)=3 (2 r+3) \log \left(\frac{1}{r^2}\right) r^5\nonumber\\
&&+3 (r (r+1) (2 r+1)+6) r+6 (3 r+2) \log (\bar{\mu}_T)\nonumber\\
&&+3 \left(2 r^6+3 r^5-3 r-2\right) \log
   \left(1-r^2\right)+13,\label{Eq:EOMSLECs}
\end{eqnarray}
depend only on the renormalization scale $\mu$ whenever the chiral-limit mass ratio $M_{B0}/M_{D0}$ is fixed.

Explicit analytical forms of the EOMS-regularized loop functions follow doing the Feynman-parameter integrals in Eqs. (\ref{Eq:HBb})-(\ref{Eq:HTc})  and applying the redefinitions of Eqs. (\ref{Eq:EOMSLECs}). One can obtain the heavy-baryon results taking $M_{D0}=M_{B0}+\delta$ with $\delta\sim\mathcal{O}(p)$ and expanding about $M_{B0}\sim \infty$
\begin{eqnarray}
&&H^{(b)}_B\simeq H^{(c)}_D\simeq -\pi\,m^3,\nonumber\\
&&H^{(c)}_B\simeq -\left(\delta^2-\frac{3}{2}m^2\right)\delta\log{\frac{m^2}{4\delta^2}}-\frac{m^2\delta}{2}-2W(m,\delta),\nonumber\\
&&H^{(b)}_T\simeq \left(\delta^2-\frac{3}{2}m^2\right)\delta\log{\frac{m^2}{4\delta^2}}+\frac{m^2\delta}{2}-2W(m,-\delta),\nonumber\\
\end{eqnarray}
with the function $W(m,\delta)$ defined as
\begin{eqnarray}
W(m,\delta)=\left\{\begin{array}{c}
(m^2-\delta^2)^{\frac{3}{2}}\arccos\frac{\delta}{m}\hspace{2.cm}m\geq|\,\delta\,| \nonumber\\
(\delta^2-m^2)^{\frac{3}{2}}\log\left(\frac{\delta}{m}+\sqrt{\frac{\delta^2}{m^2}-1}\right)  \hspace{0.1cm}m<|\,\delta\,|
            \end{array}\right..
\end{eqnarray}

\end{document}